\documentclass[runningheads]{llncs}

\usepackage[T1]{fontenc}
\usepackage{cite}
\usepackage{amsmath,amssymb,amsfonts}
\usepackage{algorithmic}
\usepackage{graphicx}
\usepackage{textcomp}
\usepackage{xcolor}
\usepackage{orcidlink}
\usepackage{booktabs}
\usepackage{tabularx}
\usepackage{bbding}
\usepackage[version=4]{mhchem}
\def\BibTeX{{\rm B\kern-.05em{\sc i\kern-.025em b}\kern-.08em
    T\kern-.1667em\lower.7ex\hbox{E}\kern-.125emX}}

\hypersetup{pdfborder={0 0 0}}

\usepackage{varioref}
\usepackage{hyperref}
\usepackage[capitalise]{cleveref}
    
\begin{document}

\title{Software Engineering for AI-driven Building Operation}

\author{Philipp Zech\inst{1}\orcidID{0000-0002-4952-4337}\Envelope \and
Sascha Hammes\inst{1}\orcidID{0000-0001-5821-5053} \and
Johannes Weninger\inst{2}\orcidID{0000-0002-8307-9652} \and
Jürgen Pannosch\inst{3}\orcidID{0009-0002-3946-9239} \and
Gernot Steindl\inst{3}\orcidID{0000-0002-9035-9206}}
\authorrunning{P.~Zech et al.}

\institute{University of Innsbruck, Tyrol, Austria\\
\email{\{philipp.zech,sascha.hammes\}@uibk.ac.at} \and
Bartenbach - The Lighting Innovators, Wattens, Austria \\
\email{Johannes.Weninger@bartenbach.com} \and
University of Applied Sciences Burgenland, Austria \\
\email{\{jürgen.pannosch,gernot.steindl\}@hochschule-burgenland.at}
}

\linespread{.91}

\maketitle

\begin{abstract}
Building operations are energy-inefficient. Artificial Intelligence (AI)-driven control systems promise benefits through optimization and predictive control, but deploying them in real buildings reveals a significant software engineering (SE) challenge. SE for AI practices assume digital environments where failures mean poor user experience. Buildings are different. A bad control decision wastes energy irreversibly, violates occupant comfort, or accelerates equipment wear. Although actual safety-critical failures are rare, as real building automation systems are inherently fault-tolerant, the physical and lasting nature of even minor failures fundamentally changes SE4AI requirements. Rooted in two interdisciplinary research projects in civil engineering and computer science that target the AI-driven optimization of building operations, we identify the missing perspectives in SE4AI that currently stymie the successful deployment of AI-based systems for building operations. We further share lessons learned and best practices, and discuss broader implications for engineering AI-driven building operations and cyber-physical systems more generally. Our work proposes a foundation for SE4AI in systems where failure has physical consequences — one the research agenda below will need to validate.
\end{abstract}


\section{Introduction}
\label{sec:introduction}

Buildings account for a substantial share of global energy consumption and greenhouse gas emissions~\cite{iea2024}. Traditional Building Automation Systems (BAS) rely on handcrafted, rule-based control strategies that are commissioned once and rarely updated~\cite{zech:eab:2025}. These static approaches struggle to adapt to changing weather, shifting occupancy patterns, and equipment degradation, leading to suboptimal performance and energy waste~\cite{stoffel2023evaluation}. Artificial Intelligence (AI), especially Machine Learning (ML), offers a way forward. Techniques such as Reinforcement Learning (RL) and Model Predictive Control (MPC) can learn and deploy control strategies that adapt to the complex and stochastic nature of building environments~\cite{stoffel2023evaluation,weinberg2023review}. Using historical and live data, these systems aim to improve energy efficiency while maintaining occupant comfort~\cite{esmaeili2025safe}.

However, deploying AI in buildings is not an AI problem per se. Rather, it is a Software Engineering (SE) problem. To this end, Software Engineering for AI (SE4AI) has established principles and practices for building, deploying, and maintaining AI systems~\cite{sklavenitis2025scoping}. These practices however were shaped by purely digital environments. We argue that applying them directly to building operations is insufficient, as buildings eventually are Cyber-physical Systems (CPS) at the intersection of digital and physical worlds~\cite{rajkumar2010cyber}. Wrong decisions have irreversible physical consequences, e.g., wasted energy that cannot be recovered, comfort violations that erode occupant trust, and equipment wear that shortens system lifespan. True safety-critical failures are rare as BAS are inherently fault-tolerant through thermal inertia, safety interlocks, and regulatory safeguards, and the EU AI Act classifies building AI as low-risk~\cite{euaiact2024}. But the physical nature of even minor failures fundamentally changes what SE must provide~\cite{hoenig2024explainable}.

In digital systems, failures are recoverable using fast rollbacks to contain consequences. In buildings, a faulty control decision can shorten the lifespan of Heating Ventilation and Air Conditioning (HVAC) equipment, cause lasting discomfort, or waste energy that is simply gone~\cite{myllyaho2022misbehaviour}. Digital systems interact through well-defined APIs. Building AI must contend with thermodynamics, hardware limits, and slow feedback loops governed by thermal mass. This gap is further compounded by noisy data from legacy hardware~\cite{hauer2024integrating}, siloed workflows~\cite{hammes2024market}, and the need to earn the trust of operators who require explanations and safe manual overrides~\cite{hoenig2024explainable,eckhardt2024survey}. Commensurate with this, our paper addresses the following central question:
\begin{quote}
    How can we complement SE4AI principles and practices to effectively support the development, deployment, and operation of AI-driven systems in the cyber-physical context of building operations?
\end{quote}
To answer this question, we draw from two interdisciplinary research projects on building operations (cf.~\cref{sec:background}) and contribute the following:
\begin{itemize}
    \item Identification and categorization of the key challenges that arise when engineering AI for building operations, and distinguish them from those in purely digital applications.
    \item Lessons learned and best practices derived from trial and error and practical experience.
    \item Discussion of the broader implications for SE4AI in other cyber-physical domains.
\end{itemize}
We position this as a preliminary experience paper grounded in two ongoing projects. It is aimed primarily at researchers working on AI deployment in CPS settings, and at practitioners facing the same structural problems in buildings or adjacent physical domains. Empirical validation of the proposed practices is an explicit part of the research agenda we outline in \cref{sec:conclusion}. Instead, our work provides a structured perspective on the SE4AI requirements of AI-driven building operations as a foundation for researchers and practitioners to build more robust and effective BAS. In our work, we focus primarily on HVAC and lighting control, which account for the largest share of building energy consumption. Though building operation covers more ground than this (e.g., access control, elevator management, air quality, fluid distribution) and many of the challenges we identify apply there too, our empirical basis is HVAC and lighting, and we scope our claims accordingly. 


\section{Background and Related Work}
\label{sec:background}

Our work grows out of practical experience in two ongoing interdisciplinary research projects, both targeting the optimization of building operations through AI-driven approaches.

BOREALIS\footnote{\url{https://projekte.ffg.at/projekt/5121377}} develops adaptive BAS that learn from operational data to reduce energy consumption and improve occupant comfort. It uses RL to continuously optimize HVAC and lighting systems based on historic and real-time sensor data~\cite{esmaeili2025safe}. The hard part is not the learning but instead deploying self-learning systems in real buildings where wrong control decisions cause equipment wear, efficiency losses, or occupant discomfort. BOREALIS requires safe exploration, robust monitoring, and decision-making that facility managers can actually understand and override. In practice, this means working with a real university building where occupancy is irregular, sensor coverage is partial, and early RL experiments that looked promising in simulation revealed a larger sim-to-real gap than expected once deployed, a direct motivation for the practices in \cref{sec:practices}.

ELEVATE\footnote{\url{https://projekte.ffg.at/projekt/5138997}} uses a Digital Twin (DT) and AI-driven optimization for energy-efficient building operation, with a particular focus on indoor environmental quality. It models the complex relationships between environmental parameters, such as lighting, air quality, thermal comfort, and occupant feedback, to build predictive models that autonomously adjust BAS control parameters~\cite{hauer2024integrating,hammes2024market}. The challenge is less about the models, but more about the "messiness" underneath that comprises fragmented data sources, legacy hardware, and the gap between what works in simulation and what works in a real building. Crucially, the models are not the key challenge. Instead, getting the data to a point where training is possible is the key challenge.

Both projects epitomize the same set of problems: integrating with fragmented legacy BAS, working with sparse and noisy sensor data, bridging digital planning and physical deployment, and earning the trust of the operators who have to run these systems on a daily basis.

\subsection{Related Work}

Our work sits at the intersection of SE for Building Automation (SE4BAS), SE4AI, and AI for Building Operations (AI4BOS) . In the following we review key literature in each area.

\subsubsection{SE for Building Automation (SE4BAS)}
\label{sssec:se4build}

Most recent work positions BAS as a distributed, service-oriented (SOA) software ecosystem. Studies review SOA for energy-efficient buildings and survey data-driven BAS architectures and identify recurring gaps in scalability, interoperability, and the integration of analytics with operational control~\cite{mendoza2021towards,taboada2024smart}. End-to-end MACH (microservices, API‑first, cloud‑based, headless)-aligned architectures that support DT visualization and sensor data integration have been demonstrated~\cite{chamari2023end}, alongside event-driven BIM (Building Information Modeling)--BAS--IoT platforms~\cite{brazauskas2021data}, semantic middleware~\cite{santos2021upgrading}, and software-defined building infrastructures~\cite{tricomi2021vertical}. Reviews on BIM and DTs, however, point to a persistent gap, viz.~DT models are rarely connected directly to automation functions and dynamic operation~\cite{walczyk2024building}.

At the modeling and process level, most work focuses on semantic representations and continuous engineering. Semantic interoperability frameworks for BAS~\cite{wollschlaeger2020play}, assisted requirements engineering~\cite{mai2020achieving}, and knowledge-based services for automation component modeling~\cite{wollschlaeger2021knowledge} all aim to reduce manual effort and improve reuse across projects. Mizutani and Mayer bring DevOps thinking into BAS engineering by using semantic reasoners and Web of Things descriptions to enable continuous verification across planning and commissioning~\cite{mitzutani2021semantic}. Ramanathan and Mayer extend this further with ontologies that automate control logic selection~\cite{ramanathan2025integrating} and a physics-infused DT framework that matches DT descriptions to control-program libraries~\cite{ramanathan2024match}.

On the commissioning and testing side, Zech et al.~propose tool support for model-based BAS and DT commissioning~\cite{zech:iceis:2024,zech:eab:2025,zech2025model,zech2025ecosystem}. Santos et al.~present an automated step-response testing tool for initial and retro-commissioning~\cite{santos2021austret}. Fierro et al.~describe a simulated DT environment that lets developers prototype and test building applications before touching a real system~\cite{fierro2022notes}. Further work explores intelligent API frameworks for occupancy-based HVAC integration~\cite{adepoju2025intelligent} and vendor-agnostic BAS implementations in specific facilities~\cite{bensoudan2024adaptation,he2021design}.

\subsubsection{SE for AI Systems (SE4AI)}
\label{sssec:se4ai}

SE4AI has matured into a recognized sub-discipline with a substantial body of surveys, frameworks, and empirical studies. Systematic reviews show that most work has focused on dependability, safety, and testing, while maintenance and evolution remain underexplored, and data problems dominate in practice~\cite{martinez2022software,assres2025state,giray2021software,kumeno2019software,serban2020adoption}. A common theme is that data, models, and code must be treated as co-equal artifacts within iterative, experiment-driven life cycles. In this line, several life cycle frameworks have been proposed. These cover AI engineering organized around architecture, development, and process~\cite{bosch2021engineering} along reliability- and debt-oriented ML life cycles that span data engineering, training, testing, and continuous evaluation~\cite{santhanam2019engineering,fischer2020ai}. In addition, various MLOps pipelines that integrate data handling, model learning, and operations with explicit triggers and artifact management have been proposed~\cite{steidl2023pipeline,lwakatare2020devops,martinez2021developing}. 

More targeted work covers requirements engineering for AI systems~\cite{ahmad2023requirements,belani2019requirements}, ML integration patterns and architectural tactics~\cite{sens2024large,heiland2023design,washizaki2019studying,serban2022adapting}, and quality assessment frameworks~\cite{jabborov2023taxonomy,gezici2022systematic}. A recurring finding from empirical studies is that many SE4AI problems are not really technical, but instead socio-technical, and deeply rooted in role boundaries, communication gaps, and missing documentation~\cite{amershi2019software,lwakatare2020devops,nahar2022collaboration,nguyen2020multiple}.

\subsubsection{AI for Building Operations (AI4BOS)}
\label{sssec:ai4build}

Early AI building control worked by embedding data-driven surrogates into optimization loops. Neural network models combined with genetic algorithms achieved solid supervisory energy savings in real HVAC systems~\cite{talib2021demand}, and symbolic regression has been used to learn interpretable thermal models for MPC deployment in occupied workspaces~\cite{ozawa2023data}. A large body of work since then applies RL for supervisory control. Model-based RL frameworks that learn neural surrogates and distill MPC solutions into fast policy networks show strong sample efficiency~\cite{zhang2019building,ding2024multi}. Model-free deep RL controllers map measured conditions directly to setpoints~\cite{zhong2022end}. Safe RL with control-barrier functions tackles comfort constraint satisfaction~\cite{esmaeili2025safe}. Comparative work suggests model-based optimization can outperform RL when accurate physical models are available, though RL wins on online speed and does not depend on accurate physical models~\cite{kou2021model}.

At larger scales, hierarchical multi-agent RL has been used to coordinate HVAC and distributed energy resources across hundreds of buildings~\cite{pei2025multi}. Industrial deployments report real energy and carbon reductions across building portfolios~\cite{chinnakannan2021using,Yang_2025}. More recently, physics-informed differentiable controllers that learn coupled airflow and thermal dynamics offer a compelling alternative to black-box RL~\cite{bian2024ventilation}.

\subsection{The Missing Perspective}

Research in all three areas is moving fast, but the areas rarely talk to each other. SE4BAS has built solid architectural patterns, semantic models, and DevOps workflows for modern BAS engineering (cf.~\cref{sssec:se4build}). AI4BOS has shown that RL, MPC, and physics-informed methods can deliver real energy savings, though actual field deployments remain rare (cf.~\cref{sssec:ai4build}). SE4AI has established mature practices for ML pipelines, data versioning, and continuous deployment (cf.~\cref{sssec:se4ai}). But each field solves its own piece of the problem and largely ignores the others. The accrued gap however matters. Deploying AI in buildings is not just about better algorithms or cleaner pipelines. It means operating a CPS~\cite{rajkumar2010cyber} where failures have physical consequences, testing requires virtual replicas, and the humans in the loop are legally responsible for the outcomes. SE4AI practices were designed for purely digital environments. They do not account for thermodynamics, sensor drift, or delayed feedback loops. \cref{tab:perspectives} makes this gap concrete.
\begin{table*}[bth]
    \small
    \centering
    \caption{The missing perspectives in SE4AI for building operations. The contrasts highlight differences in emphasis, not categorical absences. Domains like autonomous driving also engage with physical models; buildings are structurally distinct due to their combination of slow feedback loops, legacy infrastructure, and continuously occupied environments.}
    \label{tab:perspectives}
    \begin{tabularx}{\textwidth}{@{}p{0.5cm}XX@{}}
        \toprule
        & \textbf{General SE4AI Perspective} & \textbf{The Missing Perspective} \\
        \midrule
        \textbf{P1} & The cost of failure is digital & The cost of failure is physical \\
        \textbf{P2} & The world is data & Physical dynamics dominate over data patterns \\
        \textbf{P3} & Testing is data-driven & Testing is simulation-driven and risky \\
        \textbf{P4} & Humans are end-users & Humans are in-the-loop stakeholders \\
        \textbf{P5} & Data is abundant but often unlabeled & Data is physical, sparse, and noisy \\
        \textbf{P6} & Time scales are fast & Time scales are slow and delayed \\
        \bottomrule
    \end{tabularx}
\end{table*}

Both BOREALIS and ELEVATE run straight into this disconnect. BOREALIS needs safe RL exploration in buildings where wrong actions have real physical consequences, yet SE4AI offers no guidance on sandboxing those consequences or rebuilding operator trust after a bad decision. ELEVATE has to stitch together fragmented BIM, BAS, and sensor data streams to train models, but existing MLOps pipelines assume clean APIs and version-controlled datasets instead of legacy systems with intermittent connectivity. Neither project can fall back on A/B testing. Both need virtual testbeds as a core infrastructure, yet SE4AI treats simulation as optional. The digital-first mindset simply breaks when physical consequences enter the loop.

The challenges and practices in this paper grew out of practical experience in BOREALIS and ELEVATE. They reflect problems that recurred across both projects and different buildings, and that were discussed with project partners including facility managers, lighting engineers, and control system integrators. We do not claim a formal empirical method. The contribution is a structured reflection on ongoing project experience, and we treat it as such.


\section{Key Challenges}
\label{sec:challenges}

In \cref{sec:background}, we identified six missing perspectives (P1 - P6) in SE4AI for building operations (cf.~\cref{tab:perspectives}). These perspectives manifest as five concrete challenges that arise not from acute safety risks, but from the physical, irreversible, and long-horizon nature of building operations. These challenges are not exclusive to buildings.

\subsection{Physical Consequences and Irreversibility: P1}
\label{sec:challenge-physical}

In digital systems, bad predictions also have real costs, e.g., missed deadlines, lost sales, degraded user trust. In buildings, failures are physically felt and structurally harder to reverse. The difference is not categorical but one of degree as consequences are more tangible, more persistent, and harder to undo. We distinguish three categories of failure: (i)~\emph{efficiency losses}, where a control decision that simultaneously heats and cools wastes energy that cannot be recovered; (ii)~\emph{comfort violations}, where flickering lighting or temperature swings cause occupant complaints, and even after correction the damage to trust persists which often leads operators to disable the system; and (iii)~\emph{equipment wear}, where aggressive actuator cycling and rapid setpoint changes cause mechanical and thermal stress that shortens equipment lifespan and raises maintenance costs.

Actual critical failures like frozen pipes or equipment fires are rare. Real BAS are inherently fault-tolerant, where thermal mass provides buffering, safety interlocks prevent dangerous states, and building codes mandate protective measures. The EU AI Act classifies building AI as low-risk for this reason~\cite{euaiact2024}. But the physical nature of even minor failures changes SE4AI requirements. Standard SE4AI assumes failures are recoverable through rollback or retraining. In buildings, wasted energy stays wasted, comfort violations stay felt, and equipment wear stays accumulated. \emph{Consequently}, every control action must be validated against physical constraints and safety bounds before deployment, not because catastrophic failure is likely, but because routine consequences are irreversible and operator trust is hard to rebuild once lost.

\subsection{Laws of Physics Over APIs: P2~\&~P6}
\label{sec:challenge-physics}

An AI model sends a temperature setpoint, but the actual outcome depends on thermal mass, air flow, solar radiation, humidity, equipment capacity, hydraulic pressure, and electrical loads, none of which the model controls directly. Feedback is delayed (thermal response takes time to settle), noisy (sensors drift), and nonlinear (heat transfer does not follow simple statistical patterns). The environment is not a data source. It is a system of coupled differential equations that cannot be fully observed or controlled. \emph{Consequently}, model training must account for physical dynamics and temporal delays, as black-box pattern matching fails when the underlying patterns are governed by real, slow, and noisy physical processes.

\subsection{No Unified Development, Testing, and Deployment Infrastructure: P3}
\label{sec:challenge-testing}

Each building is a prototype. Unlike digital systems, A/B testing is impossible as deploying two conflicting control strategies in adjacent occupied zones risks real discomfort and energy waste. Shadow deployment does not work for HVAC. Simulation exists but suffers from model mismatch, since real buildings never behave exactly like their models~\cite{bs2025_1189}. Worse, the tools for simulation (e.g., EnergyPlus), control (e.g., BACnet stacks), and ML training (e.g., PyTorch) exist in separate ecosystems with no unified workflow from BIM to AI to deployment. \emph{Consequently}, development happens in simulation, but deployment regularly fails to account for the digital-to-physical gap.

\subsection{Trust, Transparency, and Human Operators in the Loop: P4}
\label{sec:challenge-trust}

Facility managers are not end users. They are legally responsible for safety and comfort. They must understand why the AI made a decision, override it when needed, and explain failures to tenants or regulators. An operator who does not trust a system will disable it. And a disabled AI delivers zero value regardless of how well it performs in evaluation. In buildings, explainability is not a nice-to-have. It is a prerequisite for adoption. \emph{Consequently}, models must be interpretable or provide meaningful post-hoc explanations, and control interfaces must include manual overrides and well-communicated safety bounds. This extends beyond facility managers to building occupants, who are rarely AI experts. Explainability for occupants requires different techniques, e.g., simple dashboards, plain-language summaries, and visual feedback, rather than the technical post-hoc explanations suited to professionals. Designing for this audience from the start is part of making a system actually adoptable.

\subsection{Scattered and Siloed Infrastructure: P5}
\label{sec:challenge-infrastructure}

A single building might have HVAC from the 1990s running BACnet, lighting from 2010 using DALI, newer sensors on Modbus, and occupancy data in a separate facility management system where none of it is designed to talk to the others. Data is sparse (many sensors fail silently), noisy (calibration drifts over time), and inconsistent (timestamps may not align). There is no common repository, no clean API, and no version-controlled dataset waiting to be consumed by a training pipeline. \emph{Consequently}, data engineering dominates development time, and data integration becomes the real bottleneck, not the model architecture.


\section{Lessons Learned and Best Practices from the Field}
\label{sec:practices}

Drawing from BOREALIS and ELEVATE, we present our best practices that directly address the challenges from \cref{sec:challenges}. These are not theoretical recommendations, but reflect what we learned from engineering and deploying AI-driven control in real buildings.

\subsection{Simulation-First Testing}
\label{ssec:sim-first}

Testing control policies directly in live buildings is risky. A policy that behaves unexpectedly can compromise occupant comfort, waste energy, or damage equipment. A simulation-first approach reduces this risk by evaluating policies in virtual environments before deployment~\cite{lee2014survey,shahim2016economic}. Tools like EnergyPlus and Modelica-based environments allow for testing policies across a wide range of operating conditions, including seasonal variations, occupancy patterns, and fault scenarios, repeatedly and without consequence. This makes it practical to explore behavior that would take months to observe in a real building. Simulation also supports formal analysis, as a virtual testbed can be instrumented to check safety and performance properties systematically.

But the sim-to-real gap is real~\cite{bs2025_1189}. Building models are calibrated on historical data and always carry uncertainty. Thermal dynamics, occupant behavior, and equipment degradation are hard to model precisely. A policy that looks good in simulation may perform differently once deployed, especially under conditions not well represented in the model. Simulation is the right place to rule out unsafe or poor-performing policies early. But it cannot replace real-system testing, and results based solely on simulation should be treated with caution~\cite{bs2025_1189}.

\subsection{Physical AI}

Building control policies act on the physical world. Sensors measure temperature, \ce{CO2}, occupancy, and energy consumption. Actuators open valves, adjust setpoints, and switch equipment on and off. The building responds, and the policy should adapt accordingly. This feedback loop is what makes building control a form of physical AI~\cite{brooks1990elephants,pfeifer2006body}. This matters for testing and validation. A policy's behavior cannot be understood by analyzing its code alone as it emerges from the interaction with the building, which is complex, noisy, and time-varying. Occupant behavior is unpredictable, equipment degrades, and weather changes. A policy that is safe and efficient under one set of conditions may not be under another~\cite{stoffel2023evaluation,esmaeili2025safe}.

This is why simulation-first testing (cf.~\cref{ssec:sim-first}) is necessary but not sufficient. Virtual testbeds like BOPTEST~\cite{blum2021building} provide standardized environments for pre-deployment evaluation, but real-world validation cannot be skipped. The building itself is part of the system. Successful deployments, such as data center cooling, addressed this by combining simulation-based pre-screening with staged deployment and continuous monitoring~\cite{lazic2018data}. Simulate, deploy cautiously, monitor closely, and redeploy~\cite{zech:eab:2025,zech2025ecosystem,zech:iceis:2024}: this pattern is emerging as a practical standard for physical AI in buildings~\cite{fierro2022notes}.

\subsection{Data Minimalism}
\label{ssec:data}

More data is not always better. In practice, data collection in buildings is expensive, unreliable, and raises privacy concerns~\cite{gdpr2016}. The case for data minimalism starts with robustness. Sensors fail, calibration drifts, and occupancy data is patchy. A policy that requires dense, high-quality data will be brittle in the field. Designing for minimal data from the start forces robustness~\cite{stoffel2023evaluation}. However, there is also a regulatory dimension. Buildings collect data about occupants, e.g., when they arrive, where they sit, or how they move. Under GDPR, this requires justification, consent, and protection~\cite{gdpr2016}. Thus, data minimalism is not just good engineering, it is a legal requirement in many jurisdictions.

Sample efficiency matters too. Collecting large amounts of real building data is slow, costly, and can only happen during operation, when poor control already has real consequences. A policy that learns from limited observations through simulation pre-training, transfer learning, or Bayesian methods is far more deployable than one that needs months of data before it can function~\cite{blum2021building,lazic2018data,esmaeili2025safe}. The principle is simple: collect only what is needed, make the most of what is available, and design for the data that is realistically there. Yet, we acknowledge the counterargument: sensor costs are falling, and denser instrumentation enables fault detection, drift correction, and more robust models. Data minimalism is not a claim against instrumentation. It is a design posture for the realistic case. When dense, reliable data is available, building on it is fine; the principle then shifts to designing safe fallback behavior for when that data disappears.

In practice, operationalizing this means starting from the control loop and working backwards: identify the minimum sensor set needed to close it, then ask what the system should do when any of those sensors is missing or stale. For HVAC, temperature and energy consumption often cover most of what is needed. Transfer learning and simulation pre-training can reduce dependence on real building data during early phases. Safe fallback modes that default to a conservative rule-based policy when data quality drops below a threshold can make the difference between a system that degrades gracefully and one that fails silently.

\subsection{Trust and Transparency}
\label{ssec:trust}

A system that works but is not understood will not be used. Facility managers must justify decisions to building owners, and occupants must accept that an automated system is making choices about their environment. Trust requires transparency, and transparency means being able to understand why the system made a particular decision~\cite{arrieta2020explainable}. An RL agent that maps sensor readings to control actions may perform well in testing but offer no insight into its reasoning. When it fails\textemdash and it will\textemdash there is no way to diagnose why~\cite{dulac2021challenges}. A facility manager who cannot understand a control decision cannot override it confidently, explain it to a building owner, or trust the system enough to leave it running unsupervised.

Explainability methods such as feature importance, saliency maps, and surrogate models can provide post-hoc explanations~\cite{molnar2020interpretable}. But these are approximations. A better approach is to design for interpretability from the start by using simpler policy architectures where possible~\cite{esmaeili2025safe}, or building explanation modules directly into the control pipeline~\cite{verma2018programmatically}. However, there is also the regulatory dimension (cf.~\cref{ssec:data}). GDPR gives individuals the right not to be subject to purely automated decisions without explanation~\cite{gdpr2016}. The EU AI Act triggers requirements for transparency, logging, and human oversight for certain BAS~\cite{euaiact2024}. These are legal obligations, not just options. A technically superior policy that cannot explain itself will thus lose to a simpler one that can be explained.

\subsection{DTs as Central Catalysts}

Many of the challenges above can be addressed together through one architectural choice: a DT of the building~\cite{zech2025model,zech2025agile,walczyk2024building,bhatia2026role,rashidi2026systematic}. A DT is more than a simulation. It is a synchronized replica of the real system and continuously updated with live sensor data~\cite{grieves2023digital} to reflect the actual state of the building at any point in time. And because it is a model, it can be used for things that cannot safely be done to the real building.

An RL agent needs to explore to learn, but exploration in a real building risks uncomfortable or unsafe states. In a DT, the agent explores freely with no real consequences~\cite{dulac2021challenges}. Beyond exploration, the DT enables what-if simulation grounded in the actual current state of the building, which makes results more reliable than generic offline simulation. Before any policy reaches a real building, it can be validated in the DT first, reducing deployment risk and shortening commissioning cycles~\cite{kimmel2025digital,zech:iceis:2024}. In addition, the DT can also generate training data for rare scenarios, which directly addresses data scarcity (cf.~\cref{ssec:data}). And because it exposes the full internal state of the system, it supports explainability (cf.~\cref{ssec:trust}) as a decision can be replayed and inspected to make it easier to understand and justify to facility managers and building owners. The DT is not just one tool among many. It provides the environment in which safe, data-efficient, and explainable building control AI becomes practical.


\section{Discussion and Implications}
\label{sec:discussion}

Our practices are not isolated techniques. Together, they form a coherent response to a structural problem: SE4AI was built for the digital world, and buildings are not digital. This has implications beyond buildings. The challenges we identified (cf.~\cref{sec:challenges}) appear in any domain where AI must act on a physical system, e.g., manufacturing lines, water treatment plants, traffic control, or energy grids. In all of these, the same mismatch exists. SE4AI assumes fast feedback, clean data, and recoverable failures. Physical systems offer none of these. What we describe for buildings is a template for a broader discipline, viz.~SE4AI for CPS.

The DT sits at the center of this template. Without a synchronized replica of the physical system, simulation-first testing is just offline experimentation. With a DT, testing is grounded in the actual state of the real system. Without a DT, explainability is a property of the model alone. With a DT, decisions can be replayed, inspected, and justified against real system behavior. The DT does not solve the sim-to-real gap, but it narrows it by keeping the virtual and physical systems continuously synchronized~\cite{boschert2016digital}.

Data minimalism deserves recognition as a first-class engineering principle. In digital systems, data abundance is assumed and the challenge is managing it. In physical systems, data is expensive, slow to collect, privacy-sensitive, and often unreliable. Designing for minimal data from the start produces systems that are more robust, more deployable, and easier to maintain. This shift from data maximalism to data minimalism is one of the clearest differences between SE4AI for digital and physical environments.

Trust works differently in physical systems too. In web applications, a user who distrusts a recommendation simply ignores it. In a building, a facility manager who distrusts the AI disables it. The consequence is not a missed click but a system that is physically present and doing nothing useful. Earning and maintaining operator trust is a first-class engineering concern, not a UX afterthought. This shifts explainability from a regulatory checkbox to a core architectural requirement.

Finally, the regulatory landscape is catching up. GDPR already constrains how occupant data can be collected and used~\cite{gdpr2016}. The EU AI Act introduces transparency and oversight obligations that will affect how building AI systems are designed, documented, and deployed~\cite{euaiact2024}. SE4AI practices for physical systems must thus account for these obligations from the start, and not retrofit to them later.

Several threats limit the generalizability of our findings. Both projects are situated in an Austrian academic context with specific regulatory, climatic, and infrastructure conditions. The practices were derived from the authors' own work, which introduces confirmation bias. The generalization to other CPS domains is argued by analogy rather than demonstrated empirically. We name these threats not to undermine the contribution, but because doing so is consistent with framing this as the start of a research agenda rather than a finished result.


\section{Conclusion and Future Work}
\label{sec:conclusion}

SE4AI is a maturing discipline, but it was built for a purely digital world. Buildings however are physical. Deploying AI in a building means acting on a system governed by thermodynamics, not statistics. Failures are irreversible, data is sparse, operators are legally responsible, and feedback is slow. We identified this gap, named the missing perspectives, and proposed five practices that together form a foundation for SE4AI in CPS. However, this is not the end of the problem. It is the start of a research agenda:
\begin{description}
    \item[Validate the practices in the field] BOREALIS and ELEVATE provide the testbeds. Each practice needs to be deployed in real buildings and measured against concrete outcomes like operator trust, data efficiency, and time to deployment.
    \item[Close the sim-to-real gap] Building DTs always carries uncertainty. Better calibration methods are needed, along with evidence that tells practitioners when simulation-based testing is sufficient and when it is not~\cite{bs2025_1189}.
    \item[Build the tooling] The workflow from BIM to DT to AI to deployment does not yet exist as an integrated pipeline. Progress starts with the interfaces between simulation environments, control systems, and ML toolchains.
    \item[Extend to other physical domains] The structural mismatch between SE4AI assumptions and physical reality is not unique to buildings. Testing whether our practices transfer to manufacturing, energy grids, and infrastructure is thus a natural next step.
\end{description}
The path from algorithm to deployed, trusted, and effective building AI is a SE problem. It eventually should be treated as one.


\begin{credits}
\subsubsection{\ackname} The research leading to these results has received funding from the Austrian Funding Agency (FFG) under grant agreements no.~5121377, BOREALIS, and no.~5138997, ELEVATE.

\end{credits}


\bibliographystyle{splncs04}
\bibliography{references}

\end{document}